\documentclass[conference, doublecolumn]{IEEEtran}
\usepackage{epsfig,graphicx,subfigure,psfrag,amsmath,cases}
\usepackage{latexsym,amssymb,amsmath,epsfig,subfigure,algorithm,mathtools}
\usepackage{algorithmic}
\usepackage{color}
\usepackage{url}
\usepackage{cite}
\usepackage{scrtime}
\usepackage{bm}
\usepackage{epstopdf}
\IEEEoverridecommandlockouts

\begin{document}

\title{A Distributed Multi-RF Chain Hybrid mmWave Scheme for Small-cell Systems}
\author{Lou~Zhao, Jiajia~Guo, Zhiqiang~Wei, Derrick~Wing~Kwan~Ng, and Jinhong Yuan~\thanks{L. Zhao is with the School of Engineering, Macquarie University, Sydney, NSW 2109, Australia (Email: lou.zhao@mq.edu.au). Part of this work has been done when he was with School of Electrical Engineering and Telecommunications, University of New South Wales.  Z.~Wei, D.~W.~K.~Ng, and J. Yuan are with the School of Electrical Engineering and Telecommunications, University of New South Wales, Sydney, NSW 2052, Australia (Email: zhiqiang.wei@student.unsw.edu.au; w.k.ng@unsw.edu.au; j.yuan@unsw.edu.au). J. Guo is with California Partners for Advanced Transportation Technology, University of California, Berkeley, California, USA (Email: guojj.thu@gmail.com). This work was supported in part by the Australia Research Council Discovery Project under Grants DP160104566 and DP180103550. D. W. K. Ng is supported under Australian Research Council's Discovery Early Career Researcher Award (DE170100137).}}
\maketitle

\vspace{0mm}
\begin{abstract}

This paper proposes a distributed hybrid millimeter wave (mmWave) scheme to exploit the structure of a Densely Deployed Distributed (DDD) small-cell-base-stations (SBSs) system for serving multiple users in a geographic area.
Both the SBSs and the users are equipped with full access hybrid architectures with multi-antenna arrays and multiple radio frequency chains.
Unlike the conventional cellular networks where users receive data streams from their nearest BSs, the users in our proposed scheme simultaneously receive data streams from different SBSs. With appropriate design of analog beamformers, co-channel multi-data-stream interference can be mitigated and the extra spatial degrees of freedom induced by the geographic distributed SBSs are exploited for data multiplexing. Analytical and simulation results show that the proposed scheme can improve the system sum-rate considerably, especially when the number of scattering components in millimeter wave channels is limited.

\end{abstract}\vspace*{-0mm}
\vspace*{-0mm}

\section{Introduction}

The fifth-generation (5G) wireless communication networks are expected to exploit millimeter wave (mmWave) frequencies to offer a massive trunk of bandwidth to achieve ultra-high data rate communications \cite{Bjornson2016b,Wei2017}.
However, mmWave communications suffer from severe propagation loss, due to the high penetration and reflection loss.
In mmWave channels, although highly directional information beams along strongest angle-of-arrival (AoA) components can be formed to compensate for the severe path loss, the sharpened beams can only provide limited degrees of freedom (DoFs) \cite{Bjornson2016b,Zhao2017,Zhao2018}.
This phenomenon is because the number of available signal propagation paths between a conventional base station (BS) with collocated antennas and users is limited in mmWave channels.

Geographically distributed mmWave systems, which provide sufficient DoFs in mmWave channels, can be adopted for facilitating data transmission.
On the one hand, the Densely Deployed Distributed (DDD) architecture can shorten the distances between transceivers, which can compensate for the propagation loss in mmWave frequency bands.
On the other hand, it can also potentially improve coverage performance \cite{Zhou2003} and spectral efficiency performance \cite{Bjornson2016b}.
It is known that the probability of the existence of LOS components from distributed small-cell-base-stations (SBSs) to the desired user increases with the density of SBSs \cite{Bai2015b}.
Thus, DDD systems can provide a large number of LOS propagation paths between SBSs and users.
However, how to exploit these extra LOS components between different SBSs and users for achieving a higher sum-rate is not known in general and is an exciting unexplored research area in mmWave systems.

Note that hardware architectures adopted at SBSs and users, e.g. fully digital architecture and hybrid architecture, have different features for exploiting these various LOS components between different SBSs and users.
For the distributed-SBS systems, the fully digital architecture \cite{Ng2017}, where each antenna is connected with a radio frequency (RF) chain, provides high system flexibility at the cost of hardware complexity and energy consumption.
In contrast, the hybrid architecture \cite{AZhang2015,Ng2017,Wei2018}, which strikes a balance between exploiting antenna array gain and minimizing the required number of RF chains\footnote{The number of RF chains is usually bounded below by the number of users.}, is more suitable for distributed-SBS systems.
Unfortunately, adopting the hybrid architecture in distributed-SBS mmWave systems brings extra problems, e.g. channel estimation problem and analog beamforming coordination problems.
In general, a distributed-SBS system is a kind of virtual multiple-input multiple-output (MIMO) systems \cite{Ng2017}.
Conventionally, coordinated beamforming/multi-cell precoding algorithms are well-studied for virtual MIMO systems based on the fully digital architecture, e.g. \cite{Lee2012,Sun2018b,Zhou2003}.
These algorithms rely on the channel state information (CSI) sharing among different SBSs.
Thus, the spectral efficiency and coverage performance improvement come at the cost of a stringent backhaul requirement for overhead exchange\footnote{Sharing CSIs and analog beamforming matrices requires a large amount of signalling overhead. Besides, the overhead increases significantly with the increasing number of antennas equipped at SBSs.} \cite{Maamari2016} and an increasing computational complexity \cite{Ngo2017}.
However, due to the hardware constraint of practical hybrid architectures, e.g. limited numbers of RF chains \cite{AZhang2015,Ng2017}, SBSs with the hybrid architecture cannot obtain CSIs for all antennas.
Under such a hardware constraint, conventional CSI-sharing strategies for distributed-SBS systems with fully digital architecture are not applicable for hybrid distributed-SBS systems.
On the other hand, users' architecture should also be taken into account.
Most works in the literature, e.g. \cite{Zhao2017,Andrews2014,Alkhateeb2015}, assume that each user is equipped with a multiple-antenna array and a single-RF chain to simplify the performance analysis and the algorithm design.
Hence, the users in existing works, e.g. \cite{Zhao2017,Andrews2014,Alkhateeb2015}, have limited capability in receiving multiple data streams.
In fact, due to the high data rate requirement in 5G, it is expected that users should equip with multiple RF chains to receive multiple independent data streams to improve the performance of the downlink transmission, e.g. virtual reality, artificial reality, high-resolution video, and vehicle to infrastructure communications.

In this work, we propose a distributed-SBS hybrid mmWave scheme with multiple RF chains equipped at both users and SBSs.
The proposed distributed-SBS scheme is to exploit multiple paths (LOS or NLOS components) in mmWave channels between users and geographically distributed SBSs \cite{Raghavan2018}.
To effectively exploit multiple different paths from different SBSs, there is an emerging need in designing analog beamforming matrices at users.
Based on the discussion as mentioned above, it is expected that the considered distributed-SBS mmWave scheme may achieve a higher sum-rate performance than that of the conventional fully digital BS system with collocated antennas, especially in sparse mmWave channels.
In this work, we aim to verify our basic concepts via a distributed-SBS system with a simplified pure LOS mmWave channel.
Also, further simulation results confirm that the distributed-SBS system significantly outperforms the conventional fully digital BS system with collocated antennas in typical mmWave environments with limited scattering components.

Notation:
The distribution of a circularly symmetric complex Gaussian (CSCG) random vector with a mean vector $\mathbf{x}$ and a covariance matrix ${\sigma}^{2}\mathbf{I}$  is denoted by ${\cal CN}(\mathbf{x},{\sigma}^{2}\mathbf{I})$, and $\sim$ means ``distributed as". $\mathrm{det}$ denotes for determination operation and $\mathbf{I}_{{P}}$ is an $P \times P$ identity matrix.

\section{Proposed Distributed-SBS Hybrid mmWave Scheme}

\subsection{Proposed Distributed-SBS Scheme}

In the paper, we propose a multi-user distributed-SBS scheme for mmWave systems, where both SBSs and users are equipped with multiple antennas and multiple RF chains.

\begin{figure}[t]\vspace{-0mm}
\centering
\includegraphics[width=3.0in]{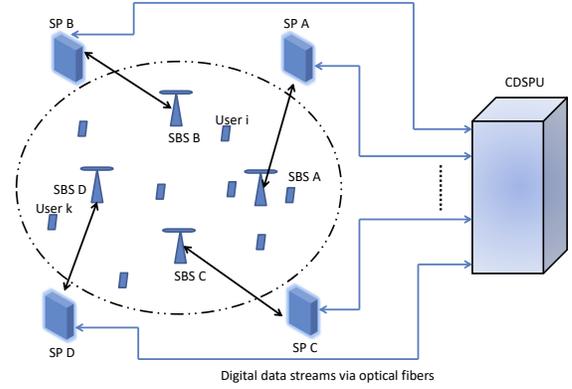}
\vspace{-2mm}
\centering
\caption{A multi-user distributed-SBS mmWave communication scheme. Here, we denote the signal processing unit as SP.}
\label{fig:MCM}\vspace{-2mm}
\end{figure}

The considered system consists of one centralized digital signal processing unit (CDSPU), $N$ SBSs, and $K$ users as shown in Fig. \ref{fig:MCM}. All $N$ SBSs are densely deployed in a small geographic area and connected to the CDSPU via optical fibers\footnote{It is reasonable to assume that the CDSPU can connect to all SBSs via optical fibers without any capacity limitation and transmission errors in a small geographic area (e.g. cell radius is less than $50$ meters).}
Each SBS is equipped with an $M$-antenna array and $N_{\mathrm{R}}$ RF chains, and it can provide at most $N_{\mathrm{R}}$ simultaneous independent data steams. Besides, SBSs and users adopt a fully access hybrid architecture where each RF chain can access an $M$-antenna uniform linear array (ULA) \cite{Zhao2017}. Also, each user is equipped with $N_{\mathrm{D}}>1$ RF chains and a $P$-antenna ULA array, where $P>N_{\mathrm{D}}$.

In a typical downlink transmission, $N$ distributed SBSs feed back estimated equivalent CSIs\footnote{The equivalent CSIs are composition of the exact mmWave channels and analog beamformers adopted at SBSs and users, as shown in Equation (\ref{DDD}) in the latter of this paper.} to the CDSPU for the design of digital precoders.
Then, the precoded transmit signals for downlink transmissions will be assigned to different SBSs from the CDSPU.
In addition, we note that distributed SBSs can split heavily loaded data traffic for the CDSPU.

In this paper, we assume $N\times N_{\mathrm{R}} = K\times N_{\mathrm{D}}$ and focus on the time-division-duplex (TDD) setting for simplicity.
With densely deployed SBSs, we assume that each SBS communicates with $K$ users simultaneously and $N_{\mathrm{D}}$ RF chains equipped at each user connect to all SBSs.
We also assume that the channel reciprocity holds. Besides, users and SBSs are fully synchronized in time and frequency.

\subsection{mmWave Channel Model}

We denote $\mathbf{H}_{i,k}\in\mathbb{C}^{M\times P}$, $ i \in \{ 1,\ldots ,N\}$ as the mmWave channel matrix between user $k$ and SBS $i$.
Besides, we assume that $\mathbf{H}_{i,k}$ is a narrowband block fading mmWave channel.
Without loss of generality, $\mathbf{H}_{i,k}\in\mathbb{C}^{M\times P}$ consists of $N_{\mathrm{cl}}$ paths and can be expressed as  \cite{Niu2015,Raghavan2018} \vspace*{-2mm}
\begin{equation}
\vspace*{-0mm}
\hspace{-0mm}\mathbf{H}_{i,k}
=\sqrt{\varpi_{{i,k}}}\sqrt{\frac{1}{N_{\mathrm{cl}}}}\overset{N_{\mathrm{cl}}}{\underset{l=1}{\sum }}{\alpha _{i,k,l}}\mathbf{h}_{i,k,l}^{\mathrm{SBS}}\mathbf{h}_{i,k,l}^{H},%
\label{Sysmodel}
\end{equation}\vspace*{-0mm}%
where $\varpi_{{i,k}}$ accounts for the corresponding average large scale path loss and $\alpha_{i,k,l}$, $l \in \{1,\cdots,N_{\mathrm{cl}}\}$, represents the complex gain for the $l$-th cluster.
We also assume that $\{\alpha_{i,k,l}\}$ are sorted in the decreasing order, i.e., $|\alpha_{i,k,1}|\geq \cdots \geq |\alpha_{i,k,N_{\mathrm{cl}}}|$. According to \cite{Hur2014}, the power ratio of the strongest cluster power over the sum of other scattered paths' power, $\varsigma_{k}=\tfrac{|\alpha_{i,k,1}|^2}{\overset{N_{\mathrm{cl}}}{\underset{l=2}{\sum }}|\alpha_{i,k,l}|^2}$, is usually larger than $1$.
Besides, $\mathbf{h}_{i,k,l}^{\mathrm{SBS}}\in\mathbb{C}^{M\times 1}$ and $\mathbf{h}_{i,k,l}\in\mathbb{C}^{P\times 1}$ are the antenna array response vectors of the SBS $i$ and user $k$ associated to the $l$-th propagation path,  which can be expressed as \vspace*{-0mm}
\begin{align}
\mathbf{h}_{i,k,l}^{\mathrm{SBS}}& =\left[
\begin{array}{ccc}
1,   \ldots
,e^{-j2\pi \left( M-1\right) \tfrac{d}{\lambda }\cos \left(
\theta _{i,k}\right) }%
\end{array}\label{Eq:4}
\right] ^{T}\text{ and} \\
\mathbf{h}_{i,k,l}& =\left[
\begin{array}{ccc}
1,  \ldots ,
e^{-j2\pi \left( P-1\right) \tfrac{d}{\lambda }\cos \left( \phi
_{i,k}\right) }%
\end{array}%
\right] ^{T}, \label{Eq:5}
\end{align}\vspace*{-0mm}%
respectively. Here, $d$ is the distance between the neighboring antennas at the BS and users and $\lambda $ is the carrier wavelength.
Variables $\theta _{i,k,l}\in \left[ 0,\text{\ }\pi \right]$ and $\phi _{i,k}\in \left[ 0,\text{\ }\pi \right] $ are the angle of incidence of the $l$-th path at antenna arrays of SBS $i$ and user $k$, respectively.

\subsection{Channel Estimation}
Here, we propose a channel estimation algorithm, which is an extension of the 5G-NR Release $14$ for above $6$ GHz bands (Tdoc R1-1611905, 2016) \cite{Giordani2018}.
For the first and second steps, we follow a similar approach as in \cite{Giordani2018}.
Then, in the third step, we propose our novel algorithm to obtain equivalent CSIs for all users at SBSs simultaneously.
First, users estimate their optimal beam directions by measuring the received power of the received CSI reference signals. Following the predefined codebook, these signals are transmitted from different SBSs simultaneously.
Second, users feed back their decided optimal beams' information to SBSs, i.e., angles-of-arrival (AoAs) of optimal beams, received energy level, and the corresponding CSI reference signals. Based on the decided optimal beam information feedback from users, SBSs can independently design their analog beamforming matrices.
After Steps $1$ and $2$, we propose to transmit orthogonal pilot symbols from users to SBSs with designed analog beamforming matrices at users and SBSs in the third step. Note that different RF chains/data-streams employ different orthogonal pilot symbols. The length of orthogonal pilot symbols equals to the number of all RF chains equipped at the users.
The uplink equivalent channels estimated at all SBSs will be fed back to the CDSPU via optical fibers. Finally, the uplink equivalent channels, $\overline{\mathbf{H}}_{\mathrm{eq}}\in\mathbb{C}^{N N_{\mathrm{R}}\times KN_{\mathrm{D}}}$, from users to SBSs can be constructed at the CDSPU and can be expressed as
\begin{align}
\hspace{-2mm}\overline{\mathbf{H}}_{\mathrm{eq}} =\hspace{-2mm} \left[\hspace{-2mm}
\begin{array}{ccc}
\left[\hspace{-2mm}
\begin{array}{c}
\mathbf{F}_{\mathrm{SBS},1}^{H}\mathbf{H}_{1,1}  \\
\vdots \\
\mathbf{F}_{\mathrm{SBS},N}^{H}\mathbf{H}_{N,1}
\end{array}\hspace{-2mm}
\right]\mathbf{F}_{1},\cdots,\left[\hspace{-2mm}
\begin{array}{c}
\mathbf{F}_{\mathrm{SBS},1}^{H}\mathbf{H}_{1,K}  \\
\vdots \\
\mathbf{F}_{\mathrm{SBS},N}^{H}\mathbf{H}_{N,K}
\end{array}\hspace{-2mm}
\right]\mathbf{F}_{K}\end{array}\hspace{-2mm}\right], \label{DDD}
\end{align}
where $\mathbf{F}_{\mathrm{SBS},j}\in\mathbb{C}^{M\times N_{\mathrm{R}}}$ is the analog beamformer for SBS $j$ and $\mathbf{F}_{k}\in\mathbb{C}^{P\times N_{\mathrm{D}}}=\left[
\begin{array}{ccc}\mathbf{f}_{k,1},\cdots,\mathbf{f}_{k,N_{\mathrm{D}}}\end{array}
\right]$, $ k \in \{1, \cdots,K \}$, is the analog beamformer for user $k$.

\section{An Illustrative Example}

In this section, we demonstrate through two basic scenarios that the proposed distributed-SBS scheme can better exploit LOS components than the conventional BS system with collocated antennas in mmWave channels. We assume pure LOS mmWave channels\footnote{There is only one LOS path between one user and one SBS, which holds for a practical indoor environment \cite{Hur2014,Halsig2017}.} and strongest AoA components are known.

\subsection{Scenario 1: Conventional BS with Collocated Antennas}

We consider $K$ $P$-antenna fully digital users and one $N\times M$-antenna fully digital BS, which usually outperforms the collocated hybrid mmWave system and serves as a benchmark for our proposed scheme. The mmWave channel matrix between user $k$ and the BS, $\mathbf{H}_{\mathrm{BS},k}\in\mathbb{C}^{NM\times P}$, can be decomposed via singular value decomposition (SVD) as:\vspace{-2mm}
\begin{align}
&\mathbf{H}_{\mathrm{BS},k}=\sqrt{\varpi_{k}}\mathbf{h}^{\mathrm{BS}}_{k}\mathbf{h}^{H}_{k}\\
&\overset{\mathrm{SVD}}{=}\left[\mathbf{U}_{\mathrm{BS},k}^\mathrm{Nzero}, \mathbf{U}_{\mathrm{BS},k}^{\mathrm{zero}} \right]\left[
\begin{array}{ccc}
\lambda_{\mathrm{BS},k} & \cdots & {0} \\
\vdots & \ddots  & \vdots \\
{0} & \cdots & 0%
\end{array}\right]\left[
\begin{array}{c}
\left(\mathbf{V}_{\mathrm{BS},k}^{\mathrm{Nzero}}\right)^{H} \\
\left(\mathbf{V}_{\mathrm{BS},k}^{\mathrm{zero}}\right)^{H}
\end{array}\right],\notag\vspace{-2mm}
\end{align}
where $\varpi_{k}$ is the path attenuation coefficient and
$\mathbf{h}^{\mathrm{BS}}_{k}$ $\in\mathbb{C}^{NM\times 1}$ and $\mathbf{h}_{k}$ $\in\mathbb{C}^{P\times 1}$ are antenna array response vectors of the BS and the user.
Besides, $\mathbf{U}_{\mathrm{BS},k}=\left[\mathbf{U}_{\mathrm{BS},k}^\mathrm{Nzero}, \mathbf{U}_{\mathrm{BS},k}^{\mathrm{zero}} \right]\in\mathbb{C}^{NM\times NM}$ and $\mathbf{V}_{\mathrm{BS},k}=\left[\mathbf{V}^\mathrm{Nzero}_{\mathrm{BS},k}, \mathbf{V}_{\mathrm{BS},k}^{\mathrm{zero}} \right]\in\mathbb{C}^{P\times P}$ are unitary matrices, where $\mathbf{U}_{\mathrm{BS},k}^{\mathrm{Nzero}}$ and $ \mathbf{U}_{\mathrm{BS},k}^{\mathrm{zero}}$ are composed of left singular vectors that correspond to nonzero singular values and zero singular values of $\mathbf{H}_{\mathrm{BS},k}$, respectively.
It is known that the DoF of a pure LOS single-user MIMO system is one, which means that the channel matrix $\mathbf{H}_{\mathrm{BS},k}$ is of rank one.
Thus, in the large number of antennas regime, the downlink achievable sum-rate of the conventional BS system by adopting the SVD algorithm, $R_{\mathrm{BS}}$, is given by
\begin{align}
\hspace{0mm}R_{\mathrm{BS}} &= \overset{K}{\underset{k=1}{\sum }}R_{\mathrm{BS},k} =\overset{K}{\underset{k=1}{\sum }}\log_2\mathrm{det}\left[\mathbf{I}+\frac{\mathbf{H}_{\mathrm{BS},k}\mathbf{H}^H_{\mathrm{BS},k}E_{\mathrm{BS},k}}{\sigma_{n}^{2}}\right]\notag\\
&\overset{NM\rightarrow\infty}{\approx} \overset{K}{\underset{k=1}{\sum }}\log_2\left[1 + \frac{{\varpi_{k}}NMP E_{\mathrm{BS},k}}{\sigma_{n}^{2}}\right]\label{C_BS},
\end{align}
where ${\lambda_{\mathrm{BS},k}\approx\sqrt{{\varpi_{k}}NMP}}$ is the approximated channel amplitude in the large number of antennas regime, $\sigma_{n}^{2}$ is the noise variance at each antenna of users and $E_{\mathrm{BS},k}$ is the transmit symbol energy for user $k$ at the BS.

\subsection{Scenario 2: Distributed SBSs}

The channel matrix between user $k$ and SBS $i$, $\mathbf{H}_{\mathrm{SBS},i,k}\in\mathbb{C}^{M\times P}$, which consists of a pure LOS component, is given by\vspace{-2mm}
\begin{align}
\mathbf{H}_{\mathrm{SBS},i,k}&=\sqrt{\varpi_{i,k}}\mathbf{h}_{i,k}^{\mathrm{SBS}}\mathbf{h}_{i,k}^{H},\vspace{-2mm}
\end{align}
where $\varpi_{i,k}$ is the path attenuation coefficient between user $k$ and SBS $i$, $\mathbf{h}_{i,k}\in\mathbb{C}^{P\times 1}$ and $\mathbf{h}^{\mathrm{SBS}}_{i,k}\in\mathbb{C}^{M\times 1}$ are the antenna array response vectors of user $k$ and SBS $i$, respectively.
Then, in the large number of antennas regime, e.g. $M$ and $P$ are sufficiently large, the channel $\mathbf{H}_{\mathrm{SBS},k}\in\mathbb{C}^{M\times P}$ can be decomposed as
\begin{align}
&\hspace{-3mm}\mathbf{H}_{\mathrm{SBS},k}=\left[\begin{array}{ccc}\sqrt{\varpi_{1,k}}\mathbf{h}_{1,k}^{\mathrm{SBS}}\mathbf{h}_{1,k}^{H}\\ \vdots \\ \sqrt{\varpi_{N,k}}\mathbf{h}_{N,k}^{\mathrm{SBS}}\mathbf{h}_{N,k}^{H}
\end{array}\right]
\overset{\mathrm{SVD}}{\approx}\left[\mathbf{U}_{\mathrm{SBS},k}^{\mathrm{Nzero}}, \mathbf{U}_{\mathrm{SBS},k}^{\mathrm{zero}} \right]\notag\\
&\hspace{-3mm}\times\left[
\begin{array}{ccccc}
\lambda_{\mathrm{SBS},1,k} & \cdots    & \cdots & {0} \\
\vdots & \ddots & \vdots &   \vdots\\
\vdots & \vdots &  \lambda_{\mathrm{SBS},N,k}  & \vdots\\                                          
{0} & \cdots &  \cdots &  {0}%
\end{array}\right]
\left[
\begin{array}{c}
\left(\mathbf{V}_{\mathrm{SBS},k}^{\mathrm{Nzero}}\right)^{H} \\
\left(\mathbf{V}_{\mathrm{SBS},k}^{\mathrm{zero}}\right)^{H}
\end{array}\right],\vspace{-2mm}
\end{align}
where $\lambda_{\mathrm{SBS},i,k}\approx\sqrt{\varpi_{i,k}M P}$ is the unique non-zero singular values of each channel matrix\footnote{When the numbers of antennas $M$ and $P$ are sufficiently large and LOS components are uniformly random, we can have a favorable propagation  \cite{Ngo2014c}. In particular, almost all singular values are concentrated around the maximum singular value $\sqrt{\varpi_{i,k}M P}$.} $\mathbf{H}_{\mathrm{SBS},i,k}$ \cite{Ngo2014c}.
The downlink achievable sum-rate of the considered distributed-SBS system by adopting the SVD algorithm can be approximated as\vspace{-1mm}
\begin{align}
R_{\mathrm{SBS}}&=\hspace{-1mm}\overset{K}{\underset{k=1}{\sum }}R_{\mathrm{SBS},k} \hspace{-1mm}=\hspace{-1mm}\overset{K}{\underset{k=1}{\sum }}\log_2\mathrm{det}\left[\mathbf{I}+\frac{\mathbf{H}_{\mathrm{SBS},k}\mathbf{H}^H_{\mathrm{SBS},k}E_{\mathrm{s},i,k}}{\sigma_{n}^{2}}\right]\notag\\
&\overset{M,P\rightarrow\infty}{\approx}\overset{K}{\underset{k=1}{\sum }} \overset{N}{\underset{i=1}{\sum }}\log_2\left[1 + \frac{{\varpi_{i,k}}MP E_{\mathrm{s},i,k}}{\sigma_{n}^{2}}\right], \label{C_SBS}\vspace{-1mm}
\end{align}
where $E_{\mathrm{s},i,k}$ is the transmit symbol energy for user $k$ at SBS $i$.
\subsection{Comparison and Discussion}

For a fair comparison, we assume that the total power constraint $P_{\mathrm{t}}$ for the two scenarios is identical and the equal power allocation strategy, i.e., $\overset{N}{\underset{i=1}{\sum }}{E_{\mathrm{s},i,k}}=NE_{\mathrm{s},k}=E_{\mathrm{BS},k}=\frac{P_{\mathrm{t}}}{K}$, where $ i \in \{1, \cdots,N \}$ and  $ k\in \{1, \cdots,K \}$ is used.
We assume that path attenuation coefficients from BS/SBSs to users are almost identical for two scenarios (e.g. indoor environment with a small cluster of users \cite{Rappaport2015}), i.e., $\varpi_{i,k}\approx\varpi_{k}\approx\varpi$, $ i \in \{1,\cdots,N\}$.
{Then, we have\vspace{-2mm}
\begin{align}
\hspace{-3mm}\Delta{R}&={R_\mathrm{SBS}}-{R_\mathrm{BS}}
\overset{M,P\rightarrow\infty}{\approx} \overset{K}{\underset{k=1}{\sum }}\log_{2}\left[ {\tfrac{{{{\left( {1 + \frac{{\mathrm{SNR}_{k}}}{N}} \right)}^N}}}{{1 + N \mathrm{SNR}_{k}}}} \right] \notag\\
&\mathop  \approx \limits^{(a)} \mathop {\mathop \sum \limits_{k = 1} }\limits^K \left( {N - 1} \right){\log _2}\left( {1 + \tfrac{{\mathrm{SNR}_{k}}}{N}} \right) - 2K{\log _2}\left( N \right), \label{C_gap}\vspace{-2mm}
\end{align}
where $\mathrm{SNR}_k \triangleq \frac{{\varpi}E_{\mathrm{BS},k}MP}{\sigma_{n}^{2}}$ denotes the received SNR at user $k$.
Note that the approximation in $(a)$ is obtained asymptotically in the high $\mathrm{SNR}_{k}$ regime, i.e., $\mathrm{SNR}_{k} \to \infty$.}

{The rate gain of the distributed-SBS scheme over the conventional BS scheme with collocated antennas versus the number of distributed SBSs $N$ is illustrated in Fig. $2$.
It is shown that the proposed distributed-SBS mmWave scheme can substantially outperform the conventional BS system by exploiting multiple LOS paths in the high SNR regime.
We also observe that the performance gap increases with an increasing number of distributed SBSs $N$.
Therefore, it is suitable to adopt the DDD-SBS architecture for communications in sparse mmWave channels.}

\begin{figure}[t]
\centering\vspace{-0mm}
\includegraphics[width=3.4in]{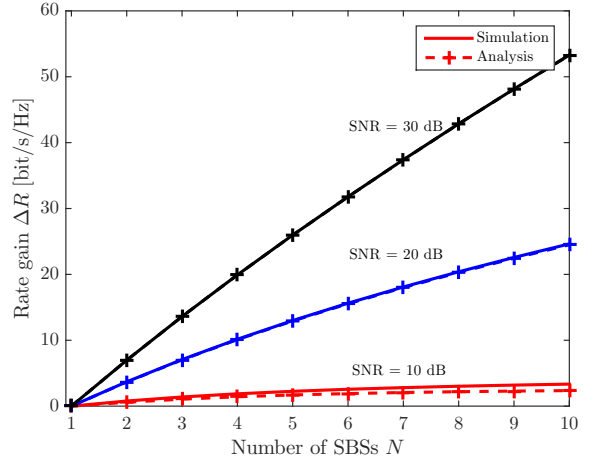}\vspace{-0mm}
\caption{The rate gain [bits/s/Hz] versus the number of SBSs.}
\label{fig:DOFGain}\vspace{-2mm}
\end{figure}
\section{Downlink Transmission with Practical Beamforming Design}
In this section, we consider a multi-user scenario where $N$ SBSs serve $K$ users.
We design effective analog beamformers for SBSs and users in this section.
An average system sum-rate, which is achieved by adopting the practical SVD digital beamforming and designed analog beamformers, serves as a performance upper bound.

\subsection{Analog Beamformers at SBSs}
Due to the inherent sparsity in mmWave channels, the number of resolvable paths between user $k$ and SBS $j$, $j \in \{1,\cdots,N\}$ is limited and the strongest path's power dominates the sum power of all paths \cite{Hur2014}.
Besides, we assume that the number of antennas $M$ equipped at SBSs is sufficiently large.
In addition, we assume that strongest propagation paths between user $k$ and all SBSs are perfectly known to user $k$ and all the SBSs.

Then, the signal transmitted from SBS $j$ for user $k$ with the designed analog beamforming vector $\bm{f}_{\mathrm{SBS},j,k}$ can be expressed as
\begin{equation}
\bm{\overline{{r}}}_{j,k} = \bm{f}_{\mathrm{SBS},j,k}{{x}}_{j,k},
\end{equation}
where $\mathrm{\mathbb{E}}[{x}_{i,k}{x}_{j,k}^{\ast}] = 0$ and $\mathrm{\mathbb{E}}[{x}_{j,k}{x}_{j,k}^{\ast}] = {E_{\mathrm{s},j,k}}$, $\forall i,k$.
We assume equal power allocation for different data streams, i.e., ${E_{\mathrm{s},i,k}} = {E_{\mathrm{s},j,k}} = \frac{E_{\mathrm{s},k}}{N}$, $\forall i,j$.
The analog beamformer, $\bm{f}_{\mathrm{SBS},j,k} $, $\forall j,k$, is designed to transmit signals via the strongest path $\mathbf{h}_{j,k,1}^{\mathrm{SBS}}$, $\forall j,k$, which can form the equivalent channel from SBS $j$ to user $k$ as
\begin{align}
&\mathbf{h}^{T}_{\mathrm{eq},j,k}=\mathbf{H}^{T}_{j,k}\bm{f}_{\mathrm{SBS},j,k} = \mathbf{H}^{T}_{j,k}\left(\tfrac{1}{\sqrt{M}}\mathbf{h}_{j,k,1}^{\mathrm{SBS}}\right)^{\ast}\notag\\
&= \sqrt{\tfrac{\varsigma_{j,k}}{\varsigma_{j,k}+1}}\mathbf{h}^{\ast}_{\mathrm{eq},j,k,1}\sqrt{M} + \sqrt{\tfrac{1}{\varsigma_{j,k}+1}}\mathbf{h}^{\ast}_{\mathrm{eq,S},j,k},  \label{Eq_per}
\end{align}
where $\mathbf{h}^{\ast}_{\mathrm{eq},j,k,1}$ is the equivalent strongest path at user $k$ from SBS $j$ and $\mathbf{h}^{\ast}_{\mathrm{eq,S},j,k}$ is the equivalent scattering component between SBS $j$ and user $k$.
In addition, the analog beamformer matrix at SBS $j$ for different users, $\mathbf{F}_{\mathrm{SBS},j}=\left[ \bm{f}_{\mathrm{SBS},j,1},\cdots,\bm{f}_{\mathrm{SBS},j,K} \right]$, satisfies the following condition in the large number of antennas regime
\begin{equation}
\mathbf{F}^{H}_{\mathrm{SBS},j}\mathbf{F}_{\mathrm{SBS},j}\overset{\mathrm{a.s.}}{\underset{M\rightarrow\infty}{=}}\mathbf{I}_{{K}},
\end{equation}
where $\mathrm{a.s.}$ stands for almost sure.
Based on Equation (\ref{Eq_per}), the single-user equivalent downlink channel for user $k$, $ {\mathbf{H}}^{T}_{\mathrm{eq},k}\in\mathbb{C}^{P \times N}$, can be given by
\begin{align}\vspace{-0mm}
\hspace{-0mm}\mathbf{H}^{T}_{\mathrm{eq},k} = \left[\mathbf{h}^{T}_{\mathrm{eq},1,k},\cdots,\mathbf{h}^{T}_{\mathrm{eq},N,k} \right].\label{ddwe}\vspace{-0mm}
\end{align}
Then, in multi-user scenarios, the signal received at the antenna array of user $k$, $\mathbf{y}_{k}$, can be expressed as\vspace{-2mm}
\begin{align}
\mathbf{y}_{k}= & \sqrt{\tfrac{M\varsigma_{j,k}}{\varsigma_{j,k}+1}} \overset{N_{\mathrm{D}} }{\underset{j=1}{\sum }}{{\mathbf{h}}}^{\ast}_{\mathrm{eq},j,k,1} {x}_{j,k}+ \sqrt{\tfrac{1}{\varsigma_{j,k}+1}}\overset{N_{\mathrm{D}} }{\underset{j=1}{\sum }}\mathbf{h}^{\ast}_{\mathrm{eq,S},j,k}{{x}}_{j,k} \notag\\
&+ \overset{N_{\mathrm{D}} }{\underset{j=1}{\sum }}\overset{K} {\underset{t=1,t\neq k}{\sum }}\mathbf{H}^{T}_{j,k}\bm{f}_{\mathrm{SBS},j,t}{{x}}_{j,t}+ \mathbf{n}, \label{EEq}\vspace{-2mm}
\end{align}
where $\mathbf{n} \sim\mathcal{CN}\left( \mathbf{0},{\sigma_{n}^{2}}\mathbf{I}_{{P}} \right)$.
\begin{figure*}[t]\setcounter{equation}{16}
\begin{align}
\overline{y}_{k,i}&\hspace{-0mm}=\hspace{-0mm}\underset{\text{Desired}\text{\ }\text{signal}}{\underbrace{\sqrt{\tfrac{M\varsigma_{i,k}}{\varsigma_{i,k}+1}}\bm{f}^{H}_{k,i}{\mathbf{h}}^{\ast}_{\mathrm{eq},i,k,1}{{x}}_{i,k}}} \hspace{-0mm}+ \hspace{-0mm}\underset{\text{Scattering\text{\ }\text{interference}}}{\underbrace{\overset{N_{\mathrm{D}} }{\underset{j=1}{\sum }}\sqrt{\tfrac{1}{\varsigma_{j,k}+1}}\bm{f}^{H}_{k,i}\mathbf{h}^{\ast}_{\mathrm{eq,S},j,k}{{x}}_{j,k}}}\hspace{-0mm}+\hspace{-0mm}\underset{\text{Co-channel\text{\ }\text{interference}}}{\underbrace{\overset{N_{\mathrm{D}}}{\underset{j=1,j\neq i}{\sum }}\sqrt{\tfrac{M\varsigma_{j,k}}{\varsigma_{j,k}+1}}\bm{f}^{H}_{k,i}{\mathbf{h}}^{\ast}_{\mathrm{eq},j,k,1}{{x}}_{ j,k}}}\notag\\
 &\hspace{-0mm}+\hspace{-0mm}\underset{\text{Inter-user\text{\ }\text{interference}}}{\underbrace{\overset{N_{\mathrm{D}} }{\underset{j=1}{\sum }}\overset{K} {\underset{t=1,t\neq k}{\sum }}\bm{f}^{H}_{k,i}\mathbf{H}^{T}_{j,t}\bm{f}_{\mathrm{SBS},j,t}{{x}}_{j,t}}}\hspace{-0mm}+ \hspace{-0mm} \underset{\text{Effective}\text{\ }\text{noise}}{\underbrace{\bm{f}^{H}_{k,i}\mathbf{n}}}. \label{EE3}
\end{align}
\hrule
\end{figure*}

\begin{figure*}[t]
\setcounter{equation}{19}
\begin{align}
&R_{\mathrm{SBS}}= \overset{K}{\underset{k=1}{\sum }}\overset{N_{\mathrm{D}}}{\underset{i=1}{\sum }}R_{k,i}
 = \notag\\
&\hspace{-0mm}\overset{K}{\underset{k=1}{\sum }}\overset{N_{\mathrm{D}}}{\underset{i=1}{\sum }} \log_2\left[1\hspace{-1mm}+\hspace{-1mm}\frac{\frac{\varsigma_{i,k}}{\varsigma_{i,k}+1}\varpi_{{i,k}}MP E_{\mathrm{s},i,k}}{{\frac{{M}}{{P}}}\hspace{-2mm}\overset{N_{\mathrm{D}}}{\underset{j=1,j\neq i}{\sum}}\hspace{-2mm}{\tfrac{\varpi_{{j,k}}\varsigma_{j,k}}{\varsigma_{j,k}+1}}|{\mathbf{h}}_{\mathrm{eq},i,k,1}^{T}{\mathbf{h}}^{\ast}_{\mathrm{eq},j,k,1}|^{2}E_{\mathrm{s},j,k}
\hspace{-1mm}+\hspace{-1mm}\overset{N_{\mathrm{D}}}{\underset{j=1}{\sum }}\hspace{-1mm}{\tfrac{\varpi_{{j,k}}}{\varsigma_{j,k}+1}}|\widehat{\bm{f}}^{H}_{k,i}\mathbf{h}^{\ast}_{\mathrm{eq,S},j,k} |^2 E_{\mathrm{s},i,k} \hspace{-1mm}+ \hspace{-1mm} \overset{N_{\mathrm{D}} }{\underset{j=1}{\sum }}\overset{K} {\underset{t=1,t\neq k}{\sum }}\hspace{-2mm}|\widehat{\bm{f}}^{H}_{k,i}\mathbf{H}^{T}_{j,t}\bm{f}_{\mathrm{SBS},j,t}|E_{\mathrm{s},j,k}\hspace{-1mm}+\hspace{-1mm} {\sigma_{n}^{2}}}\hspace{-1mm}\right]\notag\\
&\overset{(b)}{<} \hspace{-0mm}\overset{K}{\underset{k=1}{\sum }}\overset{N_{\mathrm{D}}}{\underset{i=1}{\sum }}\log_2\left[1+\hspace{-0mm}\frac{\frac{\varsigma_{i,k}}{\varsigma_{i,k}+1}\varpi_{{i,k}}MP E_{\mathrm{s},i,k}}{{\frac{{M}}{{P}}}\hspace{-0mm}\overset{N_{\mathrm{D}}}{\underset{j=1,j\neq i}{\sum}}{\tfrac{\varpi_{{j,k}}\varsigma_{j,k}}{\varsigma_{j,k}+1}}|{\mathbf{h}}_{\mathrm{eq},i,k,1}^{T}{\mathbf{h}}^{\ast}_{\mathrm{eq},j,k,1}|^{2}E_{\mathrm{s},j,k}
+ {\sigma_{n}^{2}}}\hspace{-0mm}\right].\label{SINR011}
\end{align}
\hrule
\end{figure*}
\subsection{Analog Beamformers at the User}

To design optimal analog beamformers at user $k$, we assume that CSIs are perfectly known to the CDSPU, SBSs, and user $k$.

In the single-user scenario, signals received at user $k$ after analog beamforming is given by \vspace{-0mm}
\setcounter{equation}{15}
\begin{align}
&\overline{\mathbf{y}}_{k}=\left[
\begin{array}{c}
 \mathbf{f}_{k,1} ^{H} \\
\vdots \\
 \mathbf{f}_{k,N_{\mathrm{D}}} ^{H}
\end{array}\right]\left[\mathbf{H}_{{1},k}^{T} \cdots \mathbf{H}_{{N_{\mathrm{D}}},k}^{T} \right]
\left[\begin{array}{ccc}
\mathbf{f}_{\mathrm{SBS},1,k} &\hspace{-3mm} \cdots & \hspace{-3mm}{0} \\
\vdots & \hspace{-3mm}\ddots  &\hspace{-3mm} \vdots \\
{0} & \hspace{-3mm}\cdots & \hspace{-3mm}\mathbf{f}_{\mathrm{SBS},N_{\mathrm{D}},k}%
\end{array}\right] \notag\\
&\times\left[
\begin{array}{ccc}
x_{1,k} &  &   \\
  & \hspace{-2mm}\ddots  &  \\
  &   & \hspace{-2mm} x_{N_{\mathrm{D}},k}
\end{array}\right]+\left[
\begin{array}{c}
\mathbf{f}_{k,1}^{H}\mathbf{n}   \\
\vdots   \\
\mathbf{f}_{k,N_{\mathrm{D}}}^{H}\mathbf{n}
\end{array}\right],\vspace{-0mm}
\end{align}%
where $\bm{f}_{k,i}$ is the designed analog beamformer for the $i$-th RF chain of user $k$.

Besides, in multi-user scenarios, signals received at the $i$-th RF chain of user $k$ after analog beamforming, $\overline{y}_{k,i}=\bm{f}^{H}_{k,i}\mathbf{y}_{k}$, $i \in \{1,\cdots,N_{\mathrm{D}}\}$, can be expressed as Equation (\ref{EE3}) at the top of this page.

Adopting the capacity-achieving minimum mean square error successive interference cancelation (MMSE-SIC) receiver, the analog beamformer design at the user can be formulated as an optimization problem to maximize the achievable rate of user $k$ as follows:
\setcounter{equation}{17}
\begin{align}\label{Eq_277}
\hspace{-1mm}\{\bm{f}^{\mathrm{opt}}_{k,i}\}_{i=1}^{N_{\mathrm{D}}}\hspace{-1mm} = \hspace{-1mm}
\arg\hspace{-2mm}\mathop {\max }\limits_{\{ \bm{f}_{k,i}^{{\rm{opt}}}\} _{i = 1}^{{N_{\mathrm{D}}}}}  \hspace{-2mm} \left\{ \hspace{-1mm}\log\mathrm{det} \hspace{-1mm} \left[\mathbf{I}_{{P}}\hspace{-1mm}+\hspace{-1mm}\frac{E_{\mathrm{s},k}\mathbf{F}_{k}^{H} \widetilde{\mathbf{H}}^{T}_{\mathrm{eq},k}\widetilde{\mathbf{H}}^{*}_{\mathrm{eq},k} \mathbf{F}_{k}}{N {\sigma_{n}^{2}}}\right]\hspace{-1mm}\right\},
\end{align}
where $\widetilde{\mathbf{H}}^{T}_{\mathrm{eq},k}$ is the equivalent downlink channel for user $k$ in the multi-user scenario and $\mathbf{F}_{k} = \left[ \bm{f}_{k,1}, \cdots, \bm{f}_{k,N_{\mathrm{D}}}\right]\in\mathbb{C}^{ P \times N_{\mathrm{D}}}$ collects all the analog beamformers at user $k$.
However, this optimization problem is hard to solve and only sub-optimal exhausting search algorithms exist.

For the analog beamforming matrix design, there are several practical design methods. In the following, we would like to introduce one simple but effective approache to illustrate our concepts.
The designed analog beamforming matrix for users relies on strongest AoA components\footnote{For the design of analog beamforming matrix, only strongest AoA components should be known at SBSs and users. The strongest AoA components can be obtained by following Steps $1$ and $2$ of the channel estimation procedure.} between different SBSs and users \cite{Zhao2017B} and it is given by
\begin{equation}
\widehat{\mathbf{F}}_{{k}} = \left[\widehat{\bm{f}}_{{k},1},\cdots,\widehat{\bm{f}}_{{k},N_{\mathrm{D}}}\right],\label{ddsf}
\end{equation}
where $\widehat{\bm{f}}_{{k},i}= \frac{1}{\sqrt{P}}{\mathbf{h}}_{\mathrm{eq},i,k,1}^{\ast}$ and ${\mathbf{h}}_{\mathrm{eq},i,k,1}^{\ast}$ is the strongest AoA path of ${\mathbf{h}}_{\mathrm{eq},i,k}^{\ast}$ (according to the definition of Equation (\ref{Sysmodel}) with $l=1$).
Note that, the adopted analog beamforming design in Equation (\ref{ddsf}) only requires the local CSI rather than the global CSI.

Here, we analyze the achievable sum-rate for the distributed-SBS scheme with the proposed analog beamformers.
The result is shown in Equation (\ref{SINR011}) at the top of this page.
For $(b)$ in Equation (\ref{SINR011}), we omit the impacts of scattering components as well as the inter-user interference which do not affect the system performance in the large number of antennas regime, e.g. $M\rightarrow\infty$.
The analytical results in Equation (\ref{SINR011}) illustrated that with a limited number of antennas equipped at the user, the interference caused by multiple data streams received at the same user is the bottleneck for the system sum-rate performance.
Thus, an advanced analog beamformer matrix for mitigating the inter-data-stream interference at the user is expected.

\section{Discussions and Simulations}
\begin{figure}[t]
\centering\vspace{-0mm}
\includegraphics[width=3.4in]{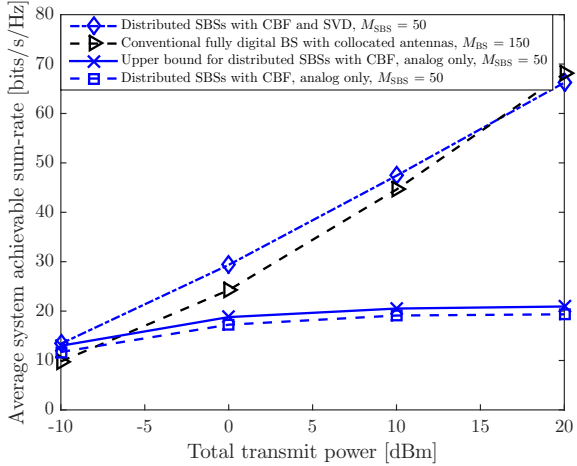}\vspace{-0mm}
\caption{An average achievable system sum-rate [bits/s/Hz] versus the total transmit power $P_{t}$ with $N=3$, $P=6$, $N_{D}=3$, $\varsigma=5$, $N_{\mathrm{cl}}=4$, and $N_{R}=K=2$.}
\label{fig:ARDB}\vspace{-2mm}
\end{figure}

\begin{figure}[t]
\centering\vspace{-0mm}
\includegraphics[width=3.4in]{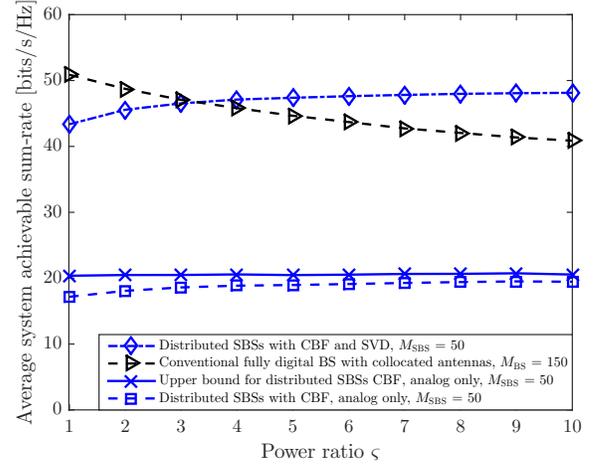}\vspace{-0mm}
\caption{An average achievable system sum-rate [bits/s/Hz] versus the power ratio $\varsigma$ with $P_{t}=10$ dBm, $N=3$, $P=6$, $N_{D}=3$, $N_{\mathrm{cl}}=4$, and $N_{\mathrm{R}}=K=2$.}
\label{fig:ARPR}\vspace{-2mm}
\end{figure}

\begin{figure}[t]
\centering\vspace{-0mm}
\includegraphics[width=3.4in]{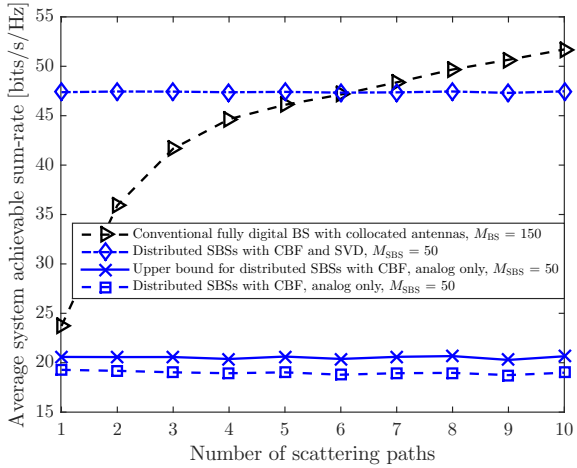}\vspace{-0mm}
\caption{An average achievable system sum-rate [bits/s/Hz] versus the number of scattering paths $N_{\mathrm{cl}}$ with $P_{t}=10$ dBm, $N=3$, $P=6$, $N_{D}=3$, $\varsigma=5$, and $N_{R}=K=2$.}
\label{fig:ARNpath}\vspace{-2mm}
\end{figure}

\begin{figure}[t]
\centering\vspace{-0mm}
\includegraphics[width=3.4in]{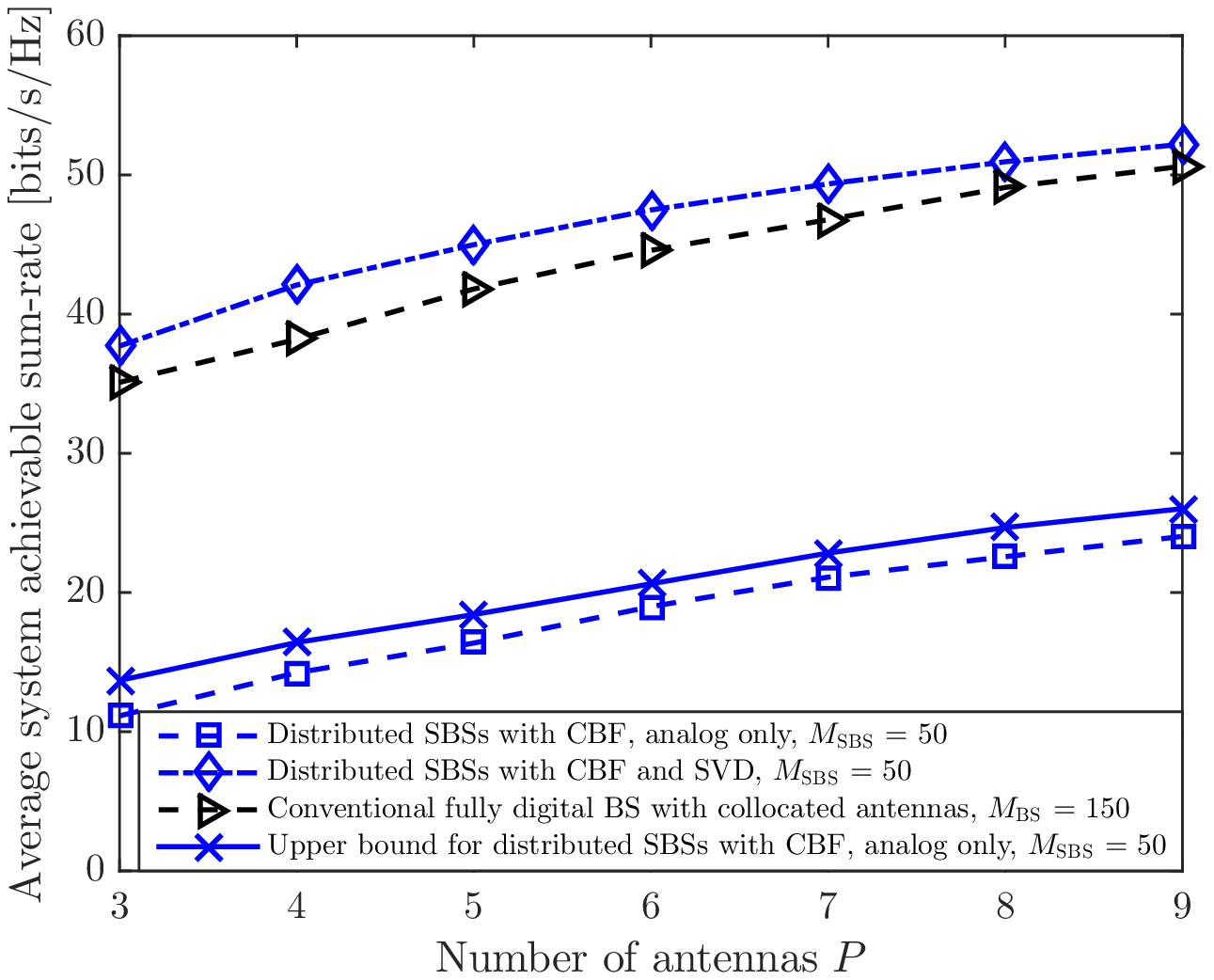}\vspace{-0mm}
\caption{An average achievable system [bits/s/Hz] versus the number of antennas $P$ with $P_{t}=10$ dBm, $\varsigma=5$, $N=3$, $N_{D}=3$, $N_{\mathrm{cl}}=4$, and $N_{\mathrm{R}}=K=2$.}
\label{fig:ARNoante}\vspace{-2mm}
\end{figure}
In this section, we illustrate some simulation results obtained from the multi-user distributed-SBS system.
To simplify the simulation, we assume that the large scale propagation path losses from SBSs to user $k$ are the same, which is reasonable in a small geographic area \cite{Rappaport2015}.
We adopt the same assumption for the conventional fully digital BS system with collocated antennas.
The simulation settings for Figs. $3$, $4$, $5$, and $6$ are as follows: the number of users $K=2$, the number of RF chains equipped at each user $N_{\mathrm{D}}=3$, the number of SBSs $N=3$, the number of antennas equipped at each SBS $M_{\mathrm{SBS}}=50$, and the number of antennas equipped at the BS $M_{\mathrm{BS}}=150$.

In Fig. \ref{fig:ARDB}, we illustrate the average achievable sum-rate versus the total transmit power.
In the low and medium SNR regimes, the distributed-SBS scheme can achieve higher sum-rate compared to the conventional fully digital BS system with collocated antennas.
However, we point out that the hardware cost and energy consumption in the distributed-SBS system are generally much smaller than the conventional fully digital BS system.

In Fig. \ref{fig:ARPR}, we illustrate the average achievable sum-rate versus the power ratio $\varsigma$. The power ratio $\varsigma$, which is defined in Equation ($1$) in Section II-B,  ranges from $1$ to $10$.
The simulation illustrates an interesting result that with an increasing power ratio $\varsigma$, the sum-rate of the conventional BS system with collocated antennas decrease and eventually only array gains can be provided, i.e., no more data multiplexing gain.
The achievable rate of the conventional BS system with collocated antennas is constrained by the number of scattering paths, which is somewhat limited in practical mmWave channels.
On the contrary, the achievable rate of the proposed distributed-SBS scheme increases with the power ratio $\varsigma$ due to the enhanced LOS components in mmWave channels.

In Fig. \ref{fig:ARNpath}, we illustrate the average achievable sum-rate versus the number of scattering paths.
With an increasing number of scattering paths, the number of non-zero eigenvalues, as well as the amplitude of eigenvalues of mmWave channels increase, which provides extra DoFs for the data transmission of the conventional fully digital BS system with collocated antennas.
Thus, in term of the achievable sum-rate, the conventional fully digital BS system with collocated antennas can outperform the distributed-SBS scheme in scattering propagation environments.
Observing Figs. \ref{fig:ARPR} and \ref{fig:ARNpath}, the rate performance gap between the distributed-SBS system and the conventional BS system with collocated antennas is considerably large when the mmWave propagation environment is sparse.

In Fig. \ref{fig:ARNoante}, we illustrate the multi-user average achievable sum-rate versus the number of antennas $P$ equipped at users.
In the simulation, we only focus on small number of antennas $P$ equipped at the user as there is only limited space at the mobile user in practice.
The simulation results show that the distributed-SBS hybrid mmWave system can outperform the conventional BS system with collocated antennas in terms of average achievable sum-rate.
These simulation results verify the key idea that the distributed-SBS system can exploit the spatial diversity of LOS components induced by the geographic distribution of SBSs in mmWave channels.

\section{Conclusions}
In this work, we proposed a distributed hybrid mmWave scheme by exploiting the features of a DDD-SBS system.
Due to the mmWave channels' inherent sparsity in the angle domain, the DoFs of the conventional fully digital BS system with collocated antennas are limited.
By distributing SBSs in a small area, the proposed scheme can exploit the spatial diversity of various paths between SBSs and users in mmWave channels.
Both analytical and simulation results have shown that significantly improvement can be obtained in terms of the achievable rate compared to the conventional BS system with collocated antennas in sparse mmWave channels.

\vspace*{-0mm}



\end{document}